\documentclass[floatfix,11pt,nofootinbib]{revtex4}
\usepackage{amssymb,amsmath}
\usepackage{graphicx,float}
\usepackage{indentfirst}

\newcommand{\beq}{\begin{equation}}
\newcommand{\eeq}{\end{equation}}
\newcommand{\bea}{\begin{eqnarray}}
\newcommand{\eea}{\end{eqnarray}}

\def\({\left(}
\def\){\right)}

\input{tcilatex}
\begin{document}

\title{Axion Electrodynamics from Infeld-van der Waerden formalisms}
\author{Andr\'{e} Martorano Kuerten}
\email{martoranokuerten@hotmail.com}
\author{A. Fernandes-Silva}
\email{armando.silva@aluno.ufabc.edu.br}
\affiliation{$^{\ast }$Independent Researcher and $^{\dagger }$Universidade Federal do
ABC,\\
09210-580, Santo Andr\'{e}, SP, Brazil}

\begin{abstract}
In this article we will explore the spacetime phase structure inside the
Infeld-van der Waerden formalisms, showing that Maxwell's theory in this
scenario leads to an axion-like phase electrodynamics, which suggests that
this phase may originate the axion field.
\end{abstract}

\maketitle

\section{Introduction}

Nowadays, the axion plays an important role in developments in several areas
of physics, such as \textrm{QCD}, in solving the strong \textrm{CP} symmetry
problem \cite{peccei,quinn}, applications in condensed matter \cite{zhang}
and string theory \cite{bakas}. The axion is also a good candidate to
explain cold dark matter \cite{preskill,abbott,fischler,gondolo}, even
though it is not directly observed in experiments. The axion/Maxwell
coupling is given by the following lagrangean term%
\begin{equation}
\mathcal{L}^{\alpha }=\frac{1}{4}\alpha F^{\mu \nu }F_{\mu \nu }^{\star }.
\end{equation}%
Since $A_{\mu }{}$ is the electromagnetic gauge potential and $e_{\mu \nu
\sigma \rho }$ are the Levi-Civita tensor components, the objects $\alpha $, 
$F_{\mu \nu }\doteqdot 2\nabla _{\lbrack \mu }A_{\nu ]}$ and $F_{\mu \nu
}^{\star }$ represent, respectively, the axion pseudo-scalar, Maxwell's
tensor and its Hodge dual defined by $F_{\mu \nu }^{\star }\doteqdot
(1/2)e_{\mu \nu }{}^{\sigma \rho }F_{\sigma \rho }$. This term is
interpreted as a topological effect in \textrm{U(1)} gauge electrodynamic
theory as it originates from a total derivative of a product between $A_{\mu
}$ and $\partial _{\mu }A_{\nu }$. In trivial topology cases the term $%
F^{\mu \nu }F_{\mu \nu }^{\star }$ must vanish. We can use the local dual
invariant electrodynamics (\textrm{LDIE}) formalism \cite%
{tiwari,tiwari2,tiwari3,tiwari4} developed by Tiwari to add the axion field
into Maxwell's electrodynamics. The \textrm{LDIE} accounts for the
electromagnetic duality symmetry by adding magnetic sources. It is also
based on Sudbery's vector lagrangean \cite{sudbery}, that considers $F_{\mu
\nu }$ as the fundamental dynamical variable. Visinelli \cite%
{visinelli,visinelli2} has also obtained a dual axion electrodynamics using
a scalar lagrangean, but the equations derived by him are not identical to
Tiwari's equations. It's worth reporting that Visinelli's equations assume
the Tiwari's pattern when a \textrm{CP}-preserving configuration is taken 
\cite{visinelli2}. An important consequence of Tiwari's works for us is the
fact that magnetic sources must be part of the equations in order to
preserve local duality invariance.

By using the $2$-spinor index formalisms \cite%
{waerden,infeld2,infeld,corson,bade}, geometric current for Infeld-van der
Waerden electromagnetic fields have been defined in \cite{kuerten,kuerten2}.
It seems that geometric source equations follow similar structures to those
obtained by Tiwari. In the $20$'s, van der Waerden \cite{waerden} provided
an electron description based on Weyl's representation and Infeld
subsequently extended Waerden's work \cite{infeld2}. Infeld together with
van-der Waerden posteriorly established the geometrical theory for curved
spacetime, thus founding the $\gamma \varepsilon $-formalisms for General
Relativity \cite{infeld}. Each $\gamma \varepsilon $ nomenclature is based
on its spinor metric, being canonical in the $\varepsilon $-formalism and
depending locally on the spacetime coordinates in the $\gamma $-formalism.
In general, the spinor decomposition of $g_{\mu \nu }$ admits a local
scale/phase freedom, such that in $\gamma $-formalism, its metric spinor
takes into account this freedom.

The Infeld-van der Waerden formalisms have been studied recently by Cardoso 
\cite{cardoso,cardoso2,cardoso3} with alternative proposals, for example 
\cite{geopho}. Sometimes classical world theories can be rewritten in its
spinor form, since the $SL(2,\mathbb{C})$ linear group of unimodular complex 
$2\times 2$ matrices has an homomorphism two to one with the orthochronous
proper Lorentz group $L_{+}^{\uparrow }$. The Infeld-van Waerden formalism
was based on the homomorphism between the Weyl and Lorentz groups. In fact,
the Weyl group $W(2,\mathbb{C})$ is longer than $SL(2,\mathbb{C})$, so the
phase invariance is expressed by 
\begin{equation}
h(e^{i\theta }g)=h(g)
\end{equation}%
with $h(g)\in SO(1,3)$ and $h(g)\in W(2,C)$, how presented in \cite{afriat}.
The correspondence between world objects living in Minkowski spacetime and
spinors living in the symplectic two-dimensional complex spinor spaces is
done using $SL(2,\mathbb{C})$ matrices. The relation between the Minkowski
metric tensor component $\eta _{\mu \nu }$ and a canonical symplectic spinor
metric component $\varepsilon _{AB}$ is given by the fundamental Clifford
algebra $\mathcal{C}\ell \left( 1,3\right) $. A remarkable fact is the
existence of a phase freedom where $\eta _{\mu \nu }$ is kept unchanged by a
gauge transformation $\varepsilon _{AB}\mapsto e^{\Theta i}\varepsilon _{AB}$%
. For curved spacetimes, the $SL(2,\mathbb{C})$ matrices are substituted by
connecting objects $\Sigma _{\mu }^{AA^{\prime }}$,\ which are generic
Hermitian matrices that depend locally on the spacetime coordinates. The
role of $SL(2,\mathbb{C})$ is played by the generalized Weyl gauge group,
which implement such transformations in General Relativity \cite{weyl}.
Thus, on each spacetime point we can connect $g_{\mu \nu }$ to $\varepsilon
_{AB}$ by using $\Sigma _{\mu }^{AA^{\prime }}$. We have the freedom to
transform the objects as follows 
\begin{equation}
g_{\mu \nu }\mapsto g_{\mu \nu }:\text{ \ \ }\varepsilon _{AB}\mapsto
\left\vert \gamma \right\vert e^{\Theta i}\varepsilon _{AB}\text{ \ \ and \
\ }\Sigma _{\mu }^{AA^{\prime }}\mapsto \left\vert \gamma \right\vert
^{-1}\Sigma _{\mu }^{AA^{\prime }},  \label{tms}
\end{equation}%
such that the metric invariance is extended by an extra scale transformation
expressed by $\left\vert \gamma \right\vert $.

Originally, this freedom has been interpreted as the geometrical origin of
the electromagnetic potential, since it would lead to an imaginary part of
the spinor connection trace $\gamma _{\mu A}{}^{A}$\ in $\gamma $-formalism,
such that it would satisfy Weyl's principle of gauge invariance \cite{weyl}.
Weyl was working on the relationship between the tetrad formalism for curved
spacetime and the parameter $\lambda $ of the Dirac $4$-spinor phase
transformation: $\Psi \mapsto e^{i\lambda }\Psi $. He noticed that if the
tetrad varies, $\lambda $ must vary too (for mathematical details, see \cite%
{weyl,afriat}). Infeld and van der Waerden, following Weyl's steps,
considered using the phase transformation of $\Psi $ to implement Dirac's
theory in curved spacetime.

In fact, given generic spin transformations: $\xi _{A}\mapsto \Lambda
_{A}{}^{B}\xi _{B}$, $\zeta ^{A}\mapsto \zeta ^{B}\Lambda _{B}^{-1}{}^{A}$
and complex conjugates, the quantity $\gamma _{\mu A}{}^{A}$ trasforms as $%
\gamma _{\mu A}{}^{A}\mapsto \gamma _{\mu A}{}^{A}+\partial _{\mu }\ln
\Delta _{\Lambda }$ \cite{infeld,bade,cardoso,cardoso2}. $\Lambda
_{A}{}^{B}(x^{\alpha })$ is some group of spin transformations for curved
spacetime and $\Delta _{\Lambda }\doteqdot \det (\Lambda _{A}{}^{B})$. In
both formalisms, the Dirac spinor is taken to be $\Psi =(\psi _{A},\chi
^{A^{\prime }})$, such that each $2$-spinor transforms as those generic
spinors. Taking into account the Weyl gauge group%
\begin{equation}
\Lambda _{B}{}^{A}=e^{i\lambda (x^{\alpha })}\delta _{A}{}^{B},  \label{wg}
\end{equation}%
useful to consider gauge invariance principle in generic backgrounds, we
have $\psi _{A}\mapsto e^{i\lambda (x^{\alpha })}\psi _{A}$, $\chi
^{A^{\prime }}\mapsto e^{i\lambda (x^{\alpha })}\chi ^{A^{\prime }}$ and $%
\Delta _{\Lambda }=e^{2i\lambda }$. Thus, the imaginary part of $\gamma
_{\mu A}{}^{A}$ transforms by the action of (\ref{wg}) as follows%
\begin{equation}
\func{Im}\gamma _{\mu A}{}^{A}\mapsto \func{Im}\gamma _{\mu
A}{}^{A}+2\partial _{\mu }\lambda .  \label{ept}
\end{equation}%
By defining the world vector $\mathcal{A}_{\mu }\doteqdot -(1/2)\func{Im}%
\gamma _{\mu A}{}^{A}$, we can see that it transforms as%
\begin{equation}
\mathcal{A}_{\mu }\mapsto \mathcal{A}_{\mu }-\partial _{\mu }\lambda ,
\end{equation}%
which is exactly how the electromagnetic potential is supposed to transform
according to the gauge invariance principle. Therefore, several authors
identified $\mathcal{A}_{\mu }$ with the electromagnetic potential vector.
This is the reason why in this scenario the literature considered the
electromagnetic fields to be \textit{intrinsic} geometrical structures,
instead of \textit{external} physical fields.

However, as remembered by Penrose \cite{honorinf}, this idea wasn't
consolidated due to physical reasons, such that the scale/phase should not
be understood on its original form. Namely, on this interpretation, the
formalism would imply a relation between electric charge and spin, since the
scale/phase couples with each type of fermion. Unfortunately, the neutron
disabled this elegant idea, as it has spin, but no electric charge.
Explicitly in \cite{cardoso3}, the equations \textrm{(5.3)} show the
couplings between Dirac spinors and $\mathcal{A}_{\mu }$, since that $\beta $%
-terms carry it.

Furthermore, the interpretation of $\mathcal{A}_{\mu }$ as an
electromagnetic potential component has impaired some investigations of
Maxwell's theory in the $\gamma $-formalism. We will work with Maxwell's
theory in the $\gamma $-formalism considering that (\ref{ept}) does not mean
an electromagnetic potential, and thus our electromagnetic fields will be
considered as \textit{external} physical entities. Therefore, we are free to
reinterpret the physical significance of the phase and scale, as well as
their possible physical consequences.

Our main motivation will be to investigate the topological origin of the
axion from $\gamma \varepsilon $-formalisms, since in different contexts the
axion is related to a topological effect. It would be interesting to
contextualize the axion electrodynamics in a spinorial structure in order to
reveal aspects of adjacent geometry and physical phenomena. We will show
that when the concept of geometrical source for Maxwell's equations in
spinorial form is applied, the mathematics suggests that it has a magnetic
nature, thus implying an \textrm{LDIE} as a consequence.

We will use the conventions $\hbar =c=1$, as well as the metric signature $%
(+---)$. The index symmetry and antisymmetry will be indicated,
respectively, by round and square brackets. The paper will be organized as
follows. In section \textrm{2} we will review the axion electrodynamics
while in the section \textrm{3}, we must present the $\gamma \varepsilon $%
-formalisms. In the section \textrm{4} we will write Maxwell's theory in
both formalisms to show that on certain aspects we can identificate $\alpha $%
\ with $\Theta $.

\section{Local Dual Axion Electrodynamics}

Originally postulated by Wilczek \cite{wilczek}, axion electrodynamics can
be taken into account by using the action $\mathcal{I}=\mathcal{I}^{f}+%
\mathcal{I}^{\alpha }+\mathcal{I}^{s}+\mathcal{I}^{k}$, where $\mathcal{I}%
^{f}$, $\mathcal{I}^{\alpha }$, $\mathcal{I}^{s}$ and $\mathcal{I}^{k}$
represent, respectively, Maxwell's , axion/Maxwell interaction, electric
source and kinetic axion actions. Explicitly, $\mathcal{I}$ is given by%
\begin{equation}
\mathcal{I}=-\frac{1}{4}\int F^{\mu \nu }F_{\mu \nu }d^{4}x+\frac{k}{4}\int
\alpha F^{\mu \nu }F_{\mu \nu }^{\star }d^{4}x+\int j_{\mu }A^{\mu }d^{4}x+%
\frac{1}{2}\int \partial _{\mu }\alpha \partial ^{\mu }\alpha d^{4}x.
\label{oae}
\end{equation}%
The object $j_{\mu }\doteqdot 4\pi \left( \rho ,-\mathbf{j}\right) $
represents the electric current density. The term $\mathcal{I}^{\alpha
}=\int \mathcal{L}^{\alpha }d^{4}x$ is the axion coupled to the \textrm{EM}
tensor, where in the 3-vector notation, we have%
\begin{equation}
\frac{1}{4}\int \alpha F^{\mu \nu }F_{\mu \nu }^{\star }d^{4}x=\int \alpha 
\mathbf{E}\bullet \mathbf{B}d^{4}x.
\end{equation}%
Taking the Euler-Lagrange equations in (\ref{oae}) with respect to $A_{\mu }$%
, we obtain%
\begin{equation}
\partial ^{\mu }F_{\mu \nu }=j_{\nu }-k\left( \partial ^{\mu }{}\alpha
\right) F_{\mu \nu }^{\star }\text{ \ \ and \ \ }\partial ^{\mu }F_{\mu \nu
}^{\star }=0.  \label{eom}
\end{equation}%
By expanding (\ref{eom}), we find the modified non-homogeneous vector
equations%
\begin{equation}
\mathbf{\nabla }\bullet \mathbf{E}+k\left( \mathbf{\nabla }\alpha \right)
\bullet \mathbf{B}=\rho ,\text{ \ \ }\mathbf{\nabla \times B}-k\left( 
\mathbf{\nabla }\alpha \right) \times \mathbf{E}-k\left( \partial _{t}\alpha
\right) \mathbf{B}-\partial _{t}\mathbf{E}=\mathbf{j}.  \label{cae}
\end{equation}%
This correction is measured by the coupling constant $k$, and its value is
estimated as being $k\sim 10^{-11}GeV^{-1}$, which is very difficult to
observe. Therefore, in the classical energy range, we must consider the
limit $k\rightarrow 0$, where the Maxwell's equations are recovered. From (%
\ref{oae}), we can take the variation with respect to the $\alpha $\ field,%
\begin{equation}
\square \alpha =\eta ^{\mu \nu }\partial _{\mu }\partial _{\nu }\alpha =k%
\mathbf{E}\bullet \mathbf{B}.
\end{equation}%
From the knowledge that $\mathbf{E}\bullet \mathbf{B}=\epsilon ^{\mu \nu
\rho \sigma }\partial _{\mu }A_{\nu }\partial _{\rho }A_{\sigma }$, we have 
\begin{equation}
\partial ^{\mu }\alpha =k\epsilon ^{\mu \nu \rho \sigma }A_{\nu }\partial
_{\rho }A_{\sigma },
\end{equation}%
which shows that the axion's influence can be interpreted as a topological
effect.

If we postulate a magnetic source $m_{\mu }\doteqdot 4\pi \left( \varrho ,-%
\mathbf{m}\right) $, Maxwell's field equations become%
\begin{equation}
\partial ^{\mu }F_{\mu \nu }=j_{\nu }\text{ \ \ \ \ and \ \ \ \ }\partial
^{\mu }F_{\mu \nu }^{\star }=m_{\nu }.  \label{eme}
\end{equation}%
The so-called electric/magnetic duality occurs when $F_{\mu \nu }\mapsto
F_{\mu \nu }^{\star }$, $F_{\mu \nu }^{\star }\mapsto -F_{\mu \nu }$, $%
j_{\mu }\mapsto m_{\mu }$ and $m_{\mu }\mapsto -j_{\mu }$ in (\ref{eme}).
Originally, this duality means that $\mathbf{E\mapsto B}$ and $\mathbf{B}%
\mapsto -\mathbf{E}$ in the vacuum, such as observed by Heaviside \cite%
{heaviside}.

We can extend the duality to a more general rotation with an arbitrary
constant rotation angle $\chi $. An extended duality transformation does not
change the mathematical structure of (\ref{eme}), and thus it does not
provide new solutions for the equations. A global duality rotation can be
represented by the field transformations%
\begin{equation}
F_{\mu \nu }\mapsto F_{\mu \nu }\cos \chi +F_{\mu \nu }^{\star }\sin \chi 
\text{ \ and \ }F_{\mu \nu }^{\star }\mapsto F_{\mu \nu }^{\star }\cos \chi
-F_{\mu \nu }\sin \chi ,  \label{dt1}
\end{equation}%
with the sources changing simultaneously as%
\begin{equation}
j_{\mu }\mapsto j_{\mu }\cos \chi +m_{\mu }\sin \chi \text{ \ and \ }m_{\mu
}\mapsto m_{\mu }\cos \chi -j_{\mu }\sin \chi .  \label{ct}
\end{equation}%
By taking the special case $\chi =\pi /2$, it is easy to see the usual
electric/magnetic duality discussed previously.

In \cite{tiwari}, Tiwari finds the axion electrodynamics equations, which we
will adapt as 
\begin{equation}
\partial ^{\mu }F_{\mu \nu }=j_{\nu }+\left( \partial ^{\mu }{}\alpha
\right) F_{\mu \nu }^{\star }\text{ \ \ \ \ and \ \ \ \ }\partial ^{\mu
}F_{\mu \nu }^{\star }=m_{\nu }-\left( \partial ^{\mu }{}\alpha \right)
F_{\mu \nu }.  \label{sudtiw}
\end{equation}%
If we consider valid the Maxwell's equations for which $m_{\mu }=0$ in (\ref%
{eom}), we obtain from (\ref{sudtiw}) the following expressions 
\begin{equation}
\left( \partial ^{\mu }{}\alpha \right) F_{\mu \nu }^{\star }=0\text{ \ \ \
\ and \ \ \ \ }\left( \partial ^{\mu }{}\alpha \right) F_{\mu \nu }=m_{\nu }.
\label{tiw}
\end{equation}%
The first relation outlines the topological effect caused by an axion field:
its presence produces a change on the eletromagnetic field components. The
last one can be interpreted as a mechanism which would impair the magnetic
monopole observation, so that the axion would cancel its effects. In
general, expressions (\ref{eme}) are invariant when (\ref{ct}) is taken with 
$\chi $ constant, while (\ref{tiw}) are invariant when $\chi =\chi
(x^{\alpha })$, and, simultaneously, the axion transforms as%
\begin{equation}
\partial _{\mu }\alpha \mapsto \partial _{\mu }\alpha +\partial _{\mu }\chi 
\text{ \ \ }\Leftrightarrow \text{ \ \ }\alpha \mapsto \alpha +\chi +c,
\label{at}
\end{equation}%
The eq. (\ref{at}) gives the behavior of the axion under a duality rotation,
with $c$ an integration constant.

\section{Spacetime Scale/Phase Invariance from Infeld-van der Waerden
Formalisms}

The group (algebra) $\mathcal{C}\ell \left( 1,3\right) $ for Minkowski
spacetime\ is represented in the $2$-spinor index formalism as follows%
\begin{equation}
\sigma _{\mu AA^{\prime }}\sigma _{\nu }^{AA^{\prime }}+\sigma _{\mu
}^{AA^{\prime }}\sigma _{\nu AA^{\prime }}=2\eta _{\mu \nu }.  \label{ca}
\end{equation}%
The Einstein summation convention will be adopted and each spinor indices
take either the values $0$,$1$ ($0^{\prime },1^{\prime }$). The objects $%
(\sigma _{\mu }^{AA^{\prime }})\in SL(2,\mathbb{C})$ are the normalized
Pauli matrices and for each \textquotedblleft world index\textquotedblright\
we have a $2\times 2$ hermitian matrix. The generalization of (\ref{ca}) for
curved space is given by%
\begin{equation}
g_{\mu \nu }(x^{\alpha })=\Sigma _{(\mu }^{AA^{\prime }}(x^{\alpha })\Sigma
_{\nu )AA^{\prime }}(x^{\alpha }),  \label{gca}
\end{equation}%
in which $g_{\mu \nu }$ is the metric tensor component for a generic
background, and the objects $\Sigma _{\mu }^{AA^{\prime }}\in \mathbb{C}$
are the Infeld-van der Waerden symbols \cite{infeld,cardoso,penrose}, which
are called connecting objects. The complex conjugation is denoted by $%
(\Sigma _{\mu }^{AB^{\prime }})^{\ast }=\Sigma _{\mu }^{A^{\prime }B}$. The
relation between spinors and tensors is established by using the hermitian
matrix set $\Sigma $, for example $v_{\mu }=\Sigma _{\mu }^{AA^{\prime
}}v_{AA^{\prime }}$ and $v_{AA^{\prime }}=\Sigma _{AA^{\prime }}^{\mu
}v_{\mu }$. In order for us to obtain the object $\Sigma _{\nu AA^{\prime }}$
from $\Sigma _{\nu }^{AA^{\prime }}$ it is necessary the use of the
\textquotedblleft metric\textquotedblright\ spinor to lower (or raise) the
spinor indexes: $\xi _{A}=\varepsilon _{BA}\xi ^{B}$, $\xi ^{A}=\varepsilon
^{AB}\xi _{B}$. The object $\varepsilon _{AB}$ is a skew-symmetric spinor
component. In the matrix form, we have%
\begin{equation}
(\varepsilon _{AB})=%
\begin{pmatrix}
0 & 1 \\ 
-1 & 0%
\end{pmatrix}%
.
\end{equation}%
With the use of the metric spinor, we can rewrite (\ref{gca}) as follows%
\begin{equation}
g_{\mu \nu }=\Sigma _{\mu }^{AA^{\prime }}\Sigma _{\nu }^{BB^{\prime
}}\varepsilon _{AB}\varepsilon _{A^{\prime }B^{\prime }}.  \label{mt}
\end{equation}%
A remarkable fact is that (\ref{mt}) is invariant under transformations (\ref%
{tms}).

In the pioneer paper \cite{infeld}, Infeld and van der Waerden created a
more general formalism given by%
\begin{equation}
\gamma _{AB}\doteqdot \left\vert \gamma \right\vert e^{\Theta i}\varepsilon
_{AB}\text{ \ \ and \ \ }\Upsilon _{\mu }^{AA^{\prime }}\doteqdot \left\vert
\gamma \right\vert ^{-1}\Sigma _{\mu }^{AA^{\prime }},  \label{sti}
\end{equation}%
in which $\left\vert \gamma \right\vert $ and $\Theta $ are real-valued
functions of some spacetime coordinates $x^{\mu }$. The object $\gamma _{AB}$
is the fundamental object of the $\gamma $-formalism, while $\Upsilon _{\mu
}^{AA^{\prime }}$ is its connecting object. We will continue to present only
aspects that will be directly used by us in our work. Detailed studies of
the $\gamma \varepsilon $-formalisms are given in the references \cite%
{cardoso,cardoso2,cardoso3}.

A crucial point is that the covariant derivative leads to an important
difference between the formalisms. In both formalisms the compatibility
metric $\nabla _{\alpha }g_{\mu \nu }=0$ implies that%
\begin{equation}
\nabla _{\mu }\left( \gamma _{AB}\gamma _{A^{\prime }B^{\prime }}\right)
=0=\nabla _{\mu }\left( \varepsilon _{AB}\varepsilon _{A^{\prime }B^{\prime
}}\right) ,  \label{cm}
\end{equation}%
since $\nabla _{\alpha }\Upsilon _{\mu }^{AA^{\prime }}=0=\nabla _{\alpha
}\Sigma _{\mu }^{AA^{\prime }}$ is valid for any hermitian matrix set \cite%
{cardoso}. Regardless of the formalism, the covariant derivative of some
spinors $\xi ^{A}$ and $\zeta _{A}$\ follows, respectively,%
\begin{equation}
\nabla _{\mu }\xi ^{A}=\partial _{\mu }\xi ^{A}+\omega _{\mu B}{}^{A}\xi ^{B}%
\text{ \ \ \ \ and \ \ \ \ }\nabla _{\mu }\zeta _{A}=\partial _{\mu }\zeta
_{A}-\omega _{\mu A}{}^{B}\zeta _{B},
\end{equation}%
with $\omega _{\mu B}{}^{A}=(\vartheta _{\mu B}{}^{A},\gamma _{\mu B}{}^{A})$%
, being the spinor connection associated with $M_{AB}=(\varepsilon
_{AB},\gamma _{AB})$. The object $\gamma _{\mu A}{}^{A}$ can be written as 
\cite{cardoso}%
\begin{equation}
\gamma _{\mu A}{}^{A}=\partial _{\mu }\ln \left\vert \gamma \right\vert -2i%
\mathcal{A}_{\mu },
\end{equation}%
with $\mathcal{A}_{\mu }=-(1/2)\func{Im}\gamma _{\mu A}{}^{A}$ being a real
world vector. The spinors $\xi ^{A}$ and $\zeta _{A}$ transform under the
action of the generalized Weyl gauge group, which can be expressed in the
component form as%
\begin{equation}
\Lambda _{A}{}^{B}=\sqrt{\varsigma }e^{i\lambda }\delta _{A}{}^{B},
\end{equation}%
in which $\varsigma >0$ is a real function and $\lambda $\ the gauge
parameter of the group.

Additionally, in the $\varepsilon $-formalism in (\ref{cm}) we have $\nabla
_{\mu }\varepsilon _{AB}=0=\nabla _{\mu }\varepsilon ^{AB}$, while in the $%
\gamma $-formalism the covariant derivative of $\gamma _{AB}$ and $\gamma
^{AB}$\ leads to%
\begin{equation}
\nabla _{\mu }\gamma _{AB}=i\beta _{\mu }\gamma _{AB}\text{ \ \ \ \ and \ \
\ \ }\nabla _{\mu }\gamma ^{AB}=-i\beta _{\mu }\gamma ^{AB}.  \label{eve}
\end{equation}%
The world object $\beta _{\mu }$ is defined as%
\begin{equation}
\beta _{\mu }\doteqdot \partial _{\mu }\Theta +2\mathcal{A}_{\mu }.
\label{eve2}
\end{equation}%
When the spacetime is flat, we have that $\gamma _{\mu A}{}^{B}=0$ such that 
$\gamma _{\mu A}{}^{A}=0$. As a consequence, $\mathcal{A}_{\mu }=0$ and $%
\gamma =\pm 1$\footnote{%
In general $\ln \left\vert \gamma \right\vert =const$. Here we will assume $%
\gamma =\pm 1$.}, so that $\beta _{\mu }$ is symply $\beta _{\mu }=\partial
_{\mu }\Theta $. \ The gauge behaviors of $\partial _{\mu }\Theta $ and $%
\mathcal{A}_{\mu }$ are given by%
\begin{equation}
\partial _{\mu }\Theta \mapsto \partial _{\mu }\Theta +2\partial _{\mu
}\lambda \text{ \ \ and \ \ }\mathcal{A}_{\mu }\mapsto \mathcal{A}_{\mu
}-\partial _{\mu }\lambda .
\end{equation}%
In the following, we will analyze the electromagnetic structure in spinor
spaces considering both formalisms, thus comparing the theory with a
magnetic source in the $\varepsilon $-formalism with the theory in the
absence of source in the $\gamma $-formalism.

\section{Maxwell's Theory and Axion-like Phase/Maxwell Coupling}

\subsection{Electromagnetic Fields}

The Maxwell's tensor and its Hodge dual can be rewritten in the $2$-spinor
notation as%
\begin{equation}
2F_{AA^{\prime }BB^{\prime }}=M_{AB}f_{A^{\prime }B^{\prime }}+M_{A^{\prime
}B^{\prime }}f_{AB},\text{ \ \ and \ \ }2F_{AA^{\prime }BB^{\prime }}^{\star
}=i\left( M_{AB}f_{A^{\prime }B^{\prime }}-M_{A^{\prime }B^{\prime
}}f_{AB}\right) ,  \label{mbs}
\end{equation}%
in which $f_{AB}$ is a symmetric object, $f_{AB}=f_{(AB)}\in \mathbb{C}$,
called Maxwell spinor. The decomposition (\ref{mbs}) is valid for any
bivector \cite{cardoso,penrose,carmeli}. For convenience, it is usual to
define the complex object $F_{\mu \nu }^{(\pm )}$\ as being $F_{\mu \nu
}^{(\pm )}\doteqdot F_{\mu \nu }\pm iF_{\mu \nu }^{\star }$. If we take the
spinor form of $F_{\mu \nu }^{(\pm )}$\ and then use (\ref{mbs}), we obtain%
\begin{equation}
F_{AA^{\prime }BB^{\prime }}^{(+)}=M_{A^{\prime }B^{\prime }}f_{AB}\text{ \
\ \ \ and \ \ \ \ }F_{AA^{\prime }BB^{\prime }}^{(-)}=M_{AB}f_{A^{\prime
}B^{\prime }}.  \label{+-}
\end{equation}%
If we contract both expressions in (\ref{mbs}) and (\ref{+-}) by $%
M^{A^{\prime }B^{\prime }}$ and $M^{AB}$, we find%
\begin{equation}
f_{AB}=F_{ABC^{\prime }}{}^{C^{\prime }}=\left( 1/2\right) F_{ABC^{\prime
}}^{(+)}{}^{C^{\prime }}\text{ \ \ \ and \ \ \ }f_{A^{\prime }B^{\prime
}}=F_{A^{\prime }B^{\prime }C}{}^{C}=\left( 1/2\right) F_{A^{\prime
}B^{\prime }C}^{(-)}{}^{C}.
\end{equation}%
Now, considering Minkowski spacetime and then using the $\varepsilon $%
-formalism, we have \cite{penrose}%
\begin{equation}
\left( f_{AB}\right) =%
\begin{pmatrix}
F_{x}-iF_{y} & -F_{z} \\ 
-F_{z} & -F_{x}-iF_{y}%
\end{pmatrix}%
,  \label{er}
\end{equation}%
in which it was used the Penrose complex $3$-vector $\mathbf{F}=\mathbf{E}-i%
\mathbf{B}$, with $\mathbf{E}$\ and $\mathbf{B}$\ being, respectively, the
electric and magnetic fields.

The invariants $F_{\mu \nu }F^{\mu \nu }$ and $F_{\mu \nu }^{\star }F^{\mu
\nu }$\ are rewritten in the spinor form as%
\begin{equation}
2F_{\mu \nu }F^{\mu \nu }=-2F_{\mu \nu }^{\star }\left( F^{\star }\right)
^{\mu \nu }=\func{Re}\left[ f_{AB}f^{AB}\right] \text{ \ \ and \ \ }2F_{\mu
\nu }^{\star }F^{\mu \nu }=\func{Im}\left[ f_{AB}f^{AB}\right] .  \label{inv}
\end{equation}%
Maxwell's action (\ref{oae}) in Minkowski background can be rewritten using (%
\ref{inv}), i. e.,%
\begin{equation}
\mathcal{I}_{M}^{f}=-\frac{1}{8}\int \mathcal{L}_{M}^{f}d^{4}x_{M}\text{ \ \
\ with \ \ \ }\mathcal{L}_{M}^{f}\doteqdot \func{Re}\left[
M^{AC}M^{BD}f_{AB}f_{CD}\right] .  \label{ma}
\end{equation}%
We must note the definition of the Maxwell's lagrangean in the $M$-formalism
given above. The volume element $d^{4}x_{M}$ is written in the $M$-formalism
as follows%
\begin{equation}
d^{4}x_{M}=\underset{=e_{AA^{\prime }BB^{\prime }CC^{\prime }DD^{\prime }}}{%
\underbrace{i\left( M_{AC}M_{BD}M_{A^{\prime }D^{\prime }}M_{B^{\prime
}C^{\prime }}-M_{AD}M_{BC}M_{A^{\prime }C^{\prime }}M_{B^{\prime }D^{\prime
}}\right) }}dx^{AA^{\prime }}dx^{BB^{\prime }}dx^{CC^{\prime
}}dx^{DD^{\prime }}.
\end{equation}%
The object $e_{AA^{\prime }BB^{\prime }CC^{\prime }DD^{\prime }}$ is the
spinor form of the Levi-Civita tensor component. If we take a closer look at
the equality $\gamma _{AB}\gamma _{C^{\prime }D^{\prime }}=\varepsilon
_{AB}\varepsilon _{C^{\prime }D^{\prime }}$, we can notice that it implies
that $d^{4}x_{\gamma }=d^{4}x_{\varepsilon }$, such that we will take into
consideration only the lagrangean term $\mathcal{L}_{M}^{f}$ of (\ref{ma}).

The objects (\ref{+-}) are useful to obtain Maxwell's spinor equations,
since each one of them carries $f_{AB}$ and $f_{A^{\prime }B^{\prime }}$ in
a unique way. In what follows, we will review Maxwell's equations with the
electric and magnetic sources in the $\varepsilon $-formalism. Later, we
will explore the Maxwell's theory in the $\gamma $-formalism without
magnetic sources to show a coupling between the \textrm{EM} tensor and the
spacetime phase.

\subsection{Maxwell's Theory in the $\protect\varepsilon $-formalism}

Let us then begin by rewriting Maxwell's equations in the $\varepsilon $%
-formalism. We will use the equations with electric and magnetic sources in
curved spacetime, i. e.,%
\begin{equation}
\nabla ^{\mu }F_{\mu \nu }=j_{\nu }\text{ \ \ \ \ and \ \ \ \ }\nabla ^{\mu
}F_{\mu \nu }^{\star }=m_{\nu }.  \label{temjm}
\end{equation}%
Also, we will consider the complex version of equations (\ref{temjm}), by
using $F_{\mu \nu }^{(\pm )}$, such that%
\begin{equation}
\nabla ^{\mu }F_{\mu \nu }^{(\pm )}=j_{\nu }\pm im_{\nu }.  \label{cme}
\end{equation}%
If we take (\ref{cme}) in its spinor form and later substitute (\ref{+-}),
we will have the following spinor expression%
\begin{equation}
\nabla _{A^{\prime }}^{B}f_{AB}=j_{AA^{\prime }}+im_{AA^{\prime }},
\label{me3}
\end{equation}%
with its complex conjugate. The expression (\ref{me3}) and its complex
conjugate represent Maxwell's equations with magnetic sources in the $%
\varepsilon $-formalism. Since one is the complex conjugate of the other, we
have, effectively, only one single spinor expression.

We must note that the global duality rotation (\ref{dt1})-(\ref{ct}) of (\ref%
{me3}) assumes the complex form%
\begin{equation}
f_{AB}\mapsto e^{-\chi i}f_{AB}\text{ \ \ \ and \ \ }j_{AA^{\prime
}}+im_{AA^{\prime }}\mapsto e^{-\chi i}\left( j_{AA^{\prime
}}+im_{AA^{\prime }}\right) ,  \label{sld}
\end{equation}%
with $\chi $\ constant. In the $\varepsilon $-formalism, the field
lagrangean (\ref{ma}) can be written as%
\begin{equation}
\mathcal{L}_{\varepsilon }^{f}=\func{Re}\left[ \varepsilon ^{AC}\varepsilon
^{BD}f_{AB}f_{CD}\right] .  \label{el}
\end{equation}%
By using (\ref{er}) in (\ref{el}), we can verify that%
\begin{equation}
\mathcal{L}_{\varepsilon }^{f}=2\left( \mathbf{B\bullet B}-\mathbf{E\bullet E%
}\right) ,
\end{equation}%
such that the result is in accordance with Maxwell's lagrangean. Another
important relation is the imaginary term in the $\varepsilon $-formalism, i.
e.,%
\begin{equation}
\func{Im}\left[ \varepsilon ^{AC}\varepsilon ^{BD}f_{AB}f_{CD}\right] =4%
\mathbf{E\bullet B,}  \label{el2}
\end{equation}%
which is in accordance with $F_{\mu \nu }^{\star }F^{\mu \nu }$. The
expression (\ref{me3}) will be used indirectly to define the geometric
magnetic source and to get to our axion-like phase electrodynamics.

\subsection{Maxwell's Theory in the $\protect\gamma $-formalism}

Let us then consider the electromagnetic equations in the $\gamma $%
-formalism without magnetic sources., i. e., $m_{\nu }=0$ in (\ref{temjm}).
The complex version will then yield%
\begin{equation}
\nabla ^{\mu }F_{\mu \nu }^{(\pm )}=j_{\nu }.  \label{cmm}
\end{equation}%
If we rewrite (\ref{cmm}) in its spinor form and use (\ref{mbs}), we will
obtain%
\begin{equation}
\nabla _{A^{\prime }}^{B}f_{AB}=j_{AA^{\prime }}+i\beta _{A^{\prime
}}^{B}f_{AB},  \label{fg}
\end{equation}%
and its complex conjugate. The $\beta $-term emerges due the eigenvalue
equations (\ref{eve}). Now, comparing (\ref{me3}) with (\ref{fg}), we are
inspired to consider a formal analogy which leads us to announce the $\beta $%
-term as a geometric magnetic current, i. e.,%
\begin{equation}
\beta _{A^{\prime }}^{B}f_{AB}\doteqdot m_{AA^{\prime }}.  \label{ms}
\end{equation}%
The idea of formally defining a geometrical source in the $\gamma $%
-formalism can be found in refs. \cite{kuerten,kuerten2}.

With the definition being established, we want to rewrite (\ref{ms}) in the
world notation. First, we note that $\beta _{A^{\prime }}^{B}f_{AB}=-\beta
^{BB^{\prime }}F_{AA^{\prime }BB^{\prime }}^{(+)}$ which implies that $\beta
^{\mu }F_{\mu AA^{\prime }}^{(+)}$ so that when we apply $\Upsilon _{\nu
}^{AA^{\prime }}$, we have%
\begin{equation}
\beta ^{\mu }F_{\mu \nu }^{(\pm )}=m_{\nu },  \label{se2}
\end{equation}%
in which the minus sign emerges when we take the complex conjugate of (\ref%
{ms}). Thus, expression (\ref{se2}) provides a world extension for Maxwell's
theory. The full world equations are given by (\ref{cmm}) and (\ref{se2}),
and we can decompose (\ref{se2}) as%
\begin{equation}
\beta ^{\mu }F_{\mu \nu }=m_{\nu }\text{ \ \ \ \ and \ \ \ }\beta ^{\mu
}F_{\mu \nu }^{\star }=0.  \label{amk}
\end{equation}%
Let us then write equations (\ref{amk}) in Minkowski spacetime. By recalling
that in this case $\beta _{\mu }=\partial _{\mu }\Theta $, expressions (\ref%
{amk}) will become%
\begin{equation}
\left( \partial ^{\mu }\Theta \right) F_{\mu \nu }=m_{\nu }\text{ \ \ and \
\ }\left( \partial ^{\mu }\Theta \right) F_{\mu \nu }^{\star }=0.
\label{amkt}
\end{equation}%
We must note the identical structure of (\ref{amk}) and (\ref{tiw}). Thus,
our system assumes the \textrm{LDIE} form where the axion field is
identified with the spacetime phase! Using the $3$-vector notation,
Maxwell's equations are supplemented by%
\begin{equation}
\left( \mathbf{\nabla }\Theta \right) \bullet \mathbf{E}=\varrho ,\text{ \ }%
\left( \mathbf{\nabla }\Theta \right) \bullet \mathbf{B}=0,\text{ \ }\left( 
\mathbf{\nabla }\Theta \right) \times \mathbf{E}+\left( \mathbf{\partial }%
_{t}\Theta \right) \mathbf{B}=\mathbf{0,}\text{ \ }\left( \mathbf{\nabla }%
\Theta \right) \times \mathbf{B}-\left( \mathbf{\partial }_{t}\Theta \right) 
\mathbf{E}=\mathbf{m.}
\end{equation}%
To conclude our study, we will analyze some aspects of local duality and
Maxwell lagrangean.

\subsection{Maxwell's Lagrangean and Local Duality Symmetry in Flat Spacetime%
}

We will work only on Minkowski background. We must consider also the case
without electric sources, i. e., $j_{AA^{\prime }}=0$. Consequently, we can
rewrite expression (\ref{fg}) as follows%
\begin{equation}
\varepsilon ^{BC}\partial _{CA^{\prime }}f_{AB}=i\varepsilon ^{BC}\left(
\partial _{CA^{\prime }}\Theta \right) f_{AB}.  \label{fg2}
\end{equation}%
A local duality rotation in the spinor notation is taken simply as $%
f_{AB}\mapsto e^{-i\chi \left( x^{\mu }\right) }f_{AB}$ in (\ref{sld}), with 
$\chi $ being a function of the spacetime coordinates. Applying the
transformation in (\ref{fg2}), we can observe that the invariance of (\ref%
{fg2}) is obtained if $\partial _{AA^{\prime }}\Theta $ transforms
simultaneously as%
\begin{equation}
\partial _{AA^{\prime }}\Theta \mapsto \partial _{AA^{\prime }}\Theta
-\partial _{AA^{\prime }}\chi \text{ \ \ }\Leftrightarrow \text{ \ \ }\Theta
\mapsto \Theta -\chi +\mathfrak{q},  \label{sd2}
\end{equation}%
with $\mathfrak{q}$ an integration constant. The phase transformation given
in (\ref{sd2}) implies that the metric spinor transforms under local duality
rotation as%
\begin{equation}
\gamma _{AB}\mapsto e^{i\left( \mathfrak{q}-\chi \right) }\gamma _{AB},
\label{dm}
\end{equation}%
since $\widetilde{\gamma }_{AB}=e^{i\widetilde{\Theta }}\varepsilon _{AB}$.

Let us then consider Maxwell's lagrangean, given in (\ref{ma}) written in
the $\gamma $-formalism, i. e.,%
\begin{equation}
\mathcal{L}_{\gamma }^{f}=\func{Re}\left[ \gamma ^{AC}\gamma
^{BD}f_{AB}f_{CD}\right] .  \label{abf}
\end{equation}%
Inserting (\ref{dm}) in (\ref{abf}) and taking $\widetilde{f}_{AB}$, we have%
\begin{equation}
\widetilde{\gamma }^{AC}\widetilde{\gamma }^{BD}\widetilde{f}_{AB}\widetilde{%
f}_{CD}=e^{-2\mathfrak{q}i}\gamma ^{AC}\gamma ^{BD}f_{AB}f_{CD},
\end{equation}%
such that Maxwell's lagrangean is local duality invariant when $\mathfrak{q}%
=0,\pm \pi ,\pm 2\pi ,\ldots $, or simply%
\begin{equation}
\widetilde{\mathcal{L}}_{\gamma }^{f}=\mathcal{L}_{\gamma }^{f},\text{ \ \ \
\ if \ \ \ \ }\mathfrak{q}=n\pi ,\text{ \ \ \ \ }n\in \mathbb{Z}.
\end{equation}%
Now, considering the duality rotation of $f_{AB}$ and (\ref{dm}) along with
expressions (\ref{mbs}) and (\ref{+-}), the electromagnetic spinor objects
transform as%
\begin{eqnarray}
2\widetilde{F}_{AA^{\prime }BB^{\prime }} &=&e^{i\mathfrak{q}}\gamma
_{AB}f_{A^{\prime }B^{\prime }}+e^{-i\mathfrak{q}}\gamma _{A^{\prime
}B^{\prime }}f_{AB},\text{ \ \ \ \ \ \ \ \ }\widetilde{F}_{AA^{\prime
}BB^{\prime }}^{(+)}=e^{-i\mathfrak{q}}F_{AA^{\prime }BB^{\prime }}^{(+)},
\label{so} \\
2\widetilde{F}_{AA^{\prime }BB^{\prime }}^{\star } &=&i\left( e^{i\mathfrak{q%
}}\gamma _{AB}f_{A^{\prime }B^{\prime }}-e^{-i\mathfrak{q}}\gamma
_{A^{\prime }B^{\prime }}f_{AB}\right) ,\text{ \ \ \ }\widetilde{F}%
_{AA^{\prime }BB^{\prime }}^{(-)}=e^{i\mathfrak{q}}F_{AA^{\prime }BB^{\prime
}}^{(-)}.  \notag
\end{eqnarray}%
We must note that (\ref{so}) unchanges when $\mathfrak{q}=m\pi $, with $m\in
2\mathbb{Z}$. In order to complete our analysis, we will consider the axion
lagrangean $\mathcal{L}^{\alpha }$ given by the interaction term%
\begin{equation}
\mathcal{L}^{\alpha }=\frac{1}{4}\alpha \mathbf{E}\bullet \mathbf{B}.
\label{api}
\end{equation}

The notation $\mathcal{L}^{\phi }$ in (\ref{api}) can denote an axion-like
coupling between $\mathbf{E}\bullet \mathbf{B}$ and any field $\phi $. In
the spinor form, we can rewrite (\ref{api}) in the $\varepsilon $-formalism
as follows%
\begin{equation}
\mathcal{L}_{\varepsilon }^{\alpha }=\alpha \func{Im}\left[ \varepsilon
^{AC}\varepsilon ^{BD}f_{AB}f_{CD}\right] ,  \label{saa}
\end{equation}%
in which we used (\ref{el2}). We must also note that%
\begin{equation}
\func{Re}\left[ \gamma ^{AC}\gamma ^{BD}f_{AB}f_{CD}\right] =\func{Re}\left[
\varepsilon ^{AC}\varepsilon ^{BD}f_{AB}f_{CD}\right] \cos \left( 2\Theta
\right) +\func{Im}\left[ \varepsilon ^{AC}\varepsilon ^{BD}f_{AB}f_{CD}%
\right] \sin \left( 2\Theta \right) .  \label{gl}
\end{equation}%
Now, combining (\ref{gl}) with (\ref{abf}) and (\ref{el}), expanding the
trigonometric functions and noting (\ref{saa}), we have%
\begin{equation}
\mathcal{L}_{\gamma }^{f}=\mathcal{L}_{\varepsilon }^{f}+2\mathcal{L}%
_{\varepsilon }^{\Theta }+\underset{j=1}{\overset{\infty }{\dsum }}\mathcal{L%
}_{\varepsilon }^{2j},  \label{pab}
\end{equation}%
with $\mathcal{L}_{\varepsilon }^{2j}$\ given by%
\begin{equation}
\mathcal{L}_{\varepsilon }^{2j}=(-1)^{j}\frac{\left( 2\Theta \right) ^{2j}}{%
\left( 2j\right) !}\left[ \mathcal{L}_{\varepsilon }^{f}+\frac{2}{2j+1}%
\mathcal{L}_{\varepsilon }^{\Theta }\right] .
\end{equation}%
Thereby, we can observe by $\mathcal{L}_{\varepsilon }^{\Theta }$ in (\ref%
{pab}) the axion-like coupling between the spacetime phase, with $f_{AB}$.

Once a small angle $\delta \Theta \simeq 0$ is taken, we have $\cos \left(
2\delta \Theta \right) \simeq 1$ and $\sin \left( 2\delta \Theta \right)
\simeq 2\delta \Theta $ in (\ref{gl}) or $\mathcal{L}_{\varepsilon
}^{2j}\simeq 0$ in (\ref{pab}), which together with the fact that they are
respectively even and odd functions, yields%
\begin{equation}
\mathcal{L}_{\gamma }^{f}\left( \pm 2\delta \Theta \right) \simeq \mathcal{L}%
_{\varepsilon }^{f}\pm 2\mathcal{L}_{\varepsilon }^{\delta \Theta }.
\end{equation}%
Since our argument is $2\Theta $, we must make an adjustment so that $\Theta 
$ runs effectively from $0$ to $\pi $. This result agrees with the axion
models! Physically, it means that the axion varies from normal to
topological insulator \cite{qi}.

\section{Conclusions and Outlook}

Based on Tiwari's work about \textrm{LDIE}, we have showed that, from the
Infeld-van der Waerden formalism it is possible to identify the axion field
as the geometric term $\Theta $. So, by restricting the equations for a flat
background, we derived the interaction Maxwell-axion term purely from
Maxwell's theory. \ It is interesting to note also that in the $\gamma $%
-formalism the equations are valid for more general spaces, which may allow
for the development of axion-related works in gravitational contexts, such
as the ones involving cold dark matter.

\section*{Acknowledgments}

The authors would like to thank Prof. R.~da~Rocha, R. T. Cavalcanti and M.
M. Cunha e Mello for the valuable suggestions and A. C. C. Esteves for
review. The research of AFS is supported by CAPES (Grant 1675676).

\end{document}